\documentclass[epj,twocolumn]{webofc}
\usepackage[varg]{txfonts}   
\woctitle{CGS15}
\begin{document}
\title{Neutrinos and the synthesis of heavy elements: the role of gravity}
%
%

\author{O. L. Caballero\inst{1,2,3}\fnsep\thanks{\email{lcaballe@umail.iu.edu}} \and
        R. Surman\inst{4} \and
        G. C. McLaughlin\inst{5}
}

\institute{Department of Physics, University of Guelph, Guelph, ON, Canada
\and ExtreMe Matter Institute EMMI, GSI Helmholtzzentrum f{\"ur} Schwerionenforschung GmbH, 64291 Darmstadt, Germany        
\and Institut f{\"u}r Kernphysik,
 Technische Universit{\"a}t Darmstadt, 64289 Darmstadt, Germany
 \and
           Department of Physics, University of Notre Dame, Notre Dame, IN, USA 
\and
           Department of Physics, North Carolina State University, Raleigh, NC, USA
          }

\abstract{%
  The synthesis of heavy elements in the Universe presents several challenges. 
  From one side the astrophysical site is still undetermined and on other hand the input 
  from nuclear physics requires the knowledge of properties of exotic nuclei, 
  some of them perhaps accessible in ion beam facilities. Black hole accretion disks 
  have been proposed as possible r-process sites. Analogously to Supernovae these objects 
  emit huge amounts of neutrinos. We discuss the neutrino emission from black hole accretion disks. 
  In particular we show the influence that the black hole strong gravitational field 
  has on changing the electron fraction relevant to the synthesis of elements.
}
\maketitle
\section{Introduction}
\label{intro}
Where and how heavy elements are produced in the Universe is one of the fundamental questions in science. 
Attempting to answer it requieres efforts from different fields. From one side, astronomical observations
provide information about the abundances of heavy elements, while nuclear physics brings
insight on the details of the reactions taking place among nuclei and that
result in some final abundances. The goal is to be able to emulate the thermodynamical conditions of some stellar site, 
to evolve a reaction network under those conditions, and finally to reproduce the observed abundaces. However, 
at this point, there is still controversy on the astronomical site(s), and the nuclear
properties of many of the nuclei participating in the reactions are uncertain. New experimental facilities bring hope 
in the study of nuclei far from stability (e.g. \citep{frib,fair,riken}). These studies 
will also shed light on theoretical models, needed to determine the properties of perhaps never
accesible nuclei.

Among all these different ingredients, neutrinos play a crucial role.
Via weak interactions they can drive a medium proton-rich or neutron-rich. This together with
the thermodynamical evolution of the matter determine what kind of elements are 
synthetized. 

The flux of neutrinos emitted from stellar sources such as supernovae and black hole accretion disks (two of the
suggested sites of r-process nucleosynthesis),
can be affected by different kind of physics, e.g flavor 
oscillations \citep{Malkus,Pllumbi} and coherent scattering \citep{Horowitz:pasta,Sonoda:pasta}.
In previous works \citep{GRCaballero,ni56}, we have studied the influence that gravity has on the 
emission of neutrinos and
the production of heavy elements in outflows 
emerging from black hole accretion disks. Although our studies have been focused on these particluar sites,
the effects of strong
gravitational fields on neutrino emission are important in any other enviroment where neutrinos
are copiously produced in the vecinity of massive central object. Of particular importance
is the consideration of the 3D geometry of the source, as the relativistic effects depend on the 
space-time curvature.
Below we discuss some more details of the effect of general relativity on 
neutrino fluxes and the synthesis of heavy elements in black hole accretion disks.

\section{General Relativistic Effects on Neutrino Fluxes}
\label{sec-1}
The strong gravitational field generated by a compact object changes the geometry of the space-time around it. 
This affects the flux of neutrinos observed at a certain distant from the central object.
The main effects of the gravitational field on the fluxes are the shifts of energies and the deformation of
the solid angle that the source subtends as seen by the observer $d\Omega_{ob}$. 
The latter can be determined via
the deflection of the neutrino trajectories and requieres to find their null geodesics in a given curvature.
The emitted energy $E_{em}$ and the observed energy $E_{ob}$ are related by
${E_{em}}=(1+z) E_{ob}$ where $(1+z)$ is the redshift. 

The effective neutrino flux observed at some distance $r_{ob}$ from the compact object is

\begin{equation}
\phi^{\mathit{eff}}=\frac{1}{4\pi}\int d\Omega_{ob}\times\phi_{ob}(E_{ob}).
\label{nufluxes}
\end{equation}

The starting point to perform a transformation from the fluxes in a flat geometry 
(here we call them Newtonian fluxes) to the ones in a curved one (General Relativistic fluxes)
is the conservation of the
number density in phase space \citep{Thorne}. This leads to write the observed general relativistic fluxes as

\begin{equation}
\phi^{\mathit{eff}} \propto \frac{1}{4\pi}\int d\Omega_{ob}\times\frac{E_{ob}^2}{\exp(E_{ob}(1+z)/T_{em})+1},
\label{observedflux}
\end{equation}
where have written the neutrino Fermi-Dirac distribution in terms of the temperature at the emission point $T_{em}$
(usually the known amount from numerical simulations). Both the redshift $1+z$ and and the solid angle $d\Omega_{ob}$
depend on the space-time geometry, and therefore on the details of the matter distribution of the compact object. 
\section{Reaction rates}
\label{sec-2}

A key factor in determining the type of nuclear products synthetized in a astrophyscial site (particularly
in neutrino-driven like environments) is the proton to neutron fraction or electron fraction $Y_e$.
If $Y_e>0.5$ the medium is proton-rich while if $Y_e<0.5$ it is neutron-rich.
In this kind of environments the initial thermodynamical conditions of the matter sorrunding 
the compact object are such that this is dissociated into electron, protons and neutrons.
Then, the main reactions setting the matter composition are:
\begin{equation}
e^{+} + n  \leftrightarrow  p + \bar{\nu}_{e},
\label{eq:capa}
\end{equation}
\begin{equation}
e^{-} + p  \leftrightarrow  n + \nu_{e}.
\label{eq:capb}
\end{equation}

If the flux of electron neutrinos is larger than the electron antineutrinos
the inverse reaction of equation \ref{eq:capb} will drive the matter
proton rich. Conversely, if the electron antineutrino flux is larger then the matter 
will be neutron rich.
On the other side, if both neutrino and antineutrino fluxes are
weak the forward reactions, electron capture on protons and positron captures on
neutrons, will play a more important role than neutrinos in determining the electron fraction.

The electron fraction $\tilde Y_e$ that is obtained by taking into account only the reverse reactions of eqs. \ref{eq:capa} and
\ref{eq:capb}
depends on the absoption rates of these processes.
As the neutrino fluxes, which determine the reaction rates, 
are affected by a strong gravitational field, $\tilde Y_e$ is also affected.
In terms of the observed fluxes $\phi^{\mathit{eff}}$ (eq. \ref{observedflux})
the rate of absortion of neutrinos on neutrons is
\begin{equation}
\lambda_{\nu_{e}n}=b\int^{\inf}_0 \phi^{\mathit{eff}}_{\nu_e}(E_{ob}+\Delta)^2\sqrt{1-\frac{m^2_e}{(E_{ob}+\Delta)^2}}W_M dE_{ob},
\end{equation}
and of antineutrinos on protons
\begin{equation}
\lambda_{\bar\nu_e p}=b\int^{\inf}_ {\Delta+m_e}\phi^{\mathit{eff}}_{\bar\nu_e}(E_{ob}-\Delta)^2\sqrt{1-\frac{m^2_e}{(E_{ob}-\Delta)^2}}W_{\bar M} dE_{ob},
\end{equation}
where $W_M=1+1.1E_{ob}/m_n$ and $W_{\bar M }= 1-7.1E_{ob}/m_n$ are the 
weak magnetism corrections \citep{Horowitz:wm}, $m_n$, $m_e$ are the neutron and electron masses respectively, $\Delta$ is the
neutron-proton mass difference, and $b=9.704\times 10^{-50}$ cm$^2$keV$^{-2}$.

\section{Astrophysical site: Accretion disk outflows}
\label{sec-3}

A possible scenario after the merger of two compact objects
(black hole-neutron star or neutron star-neutron star) is the formation
of an accretion disk or torus around a black hole. Given the inital conditions
of the progenitors the matter of disk
is neutron rich  and hot enough to be dissociated in nucleons. Some fraction of this matter
can be ejected in hot outflows,
presenting an interesting scenario for the synthesis of neutron-rich elements.

\begin{figure}
\centering
\includegraphics[width=\linewidth]{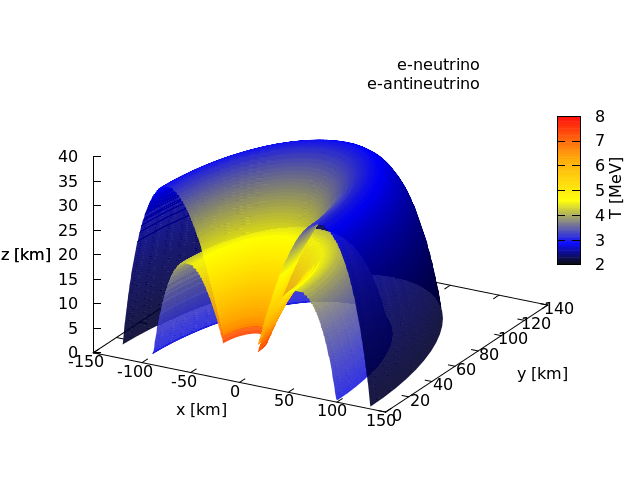}
\caption{Electron neutrino (outter) and antineutrino (inner) surfaces corresponding to a snapshop at t=20 ms, of a
hydrodynamical simulation of a torus around a 3 solar mass black hole.}
\label{fig-1}       
\end{figure}

The results presented here and on ref. \citep{ni56} are based on a time depedent hydrodynamical
simulation of accreting matter around a 3 solar masses black hole 
with a spin $a=0.8$ (for more details on the simulation see \citep{Just:2014}). In this scenario
neutrinos are coupusly emitted. The emission points correspond to the neutrino surfaces, 
the places where
after being trapped by the high density conditions, neutrinos can freely travel.
Figure \ref{fig-1} shows a transversal cut of the electron neutrino and antineutrino surfaces for
this disk model at $t=20$ ms. The $z$ axis correspond to the actual decoupling height, and
the colored scale shows the neutrino temperature $T_{em}$, which is crucial in calculating
the neutrino fluxes as described in eq. \ref{observedflux}. The reactions and details used to calculate
these surfaces are discussed in ref. \cite{Caballerosurface}. The difference in the neutrino decoupling
surfaces for each flavor, in energy and in distance from the black hole, has important consequences 
in terms of the effects of gravity on nucleosyntesis. 

\begin{figure}
\centering
\includegraphics[width=\linewidth, trim= 0mm 0mm 0mm 100mm, clip]{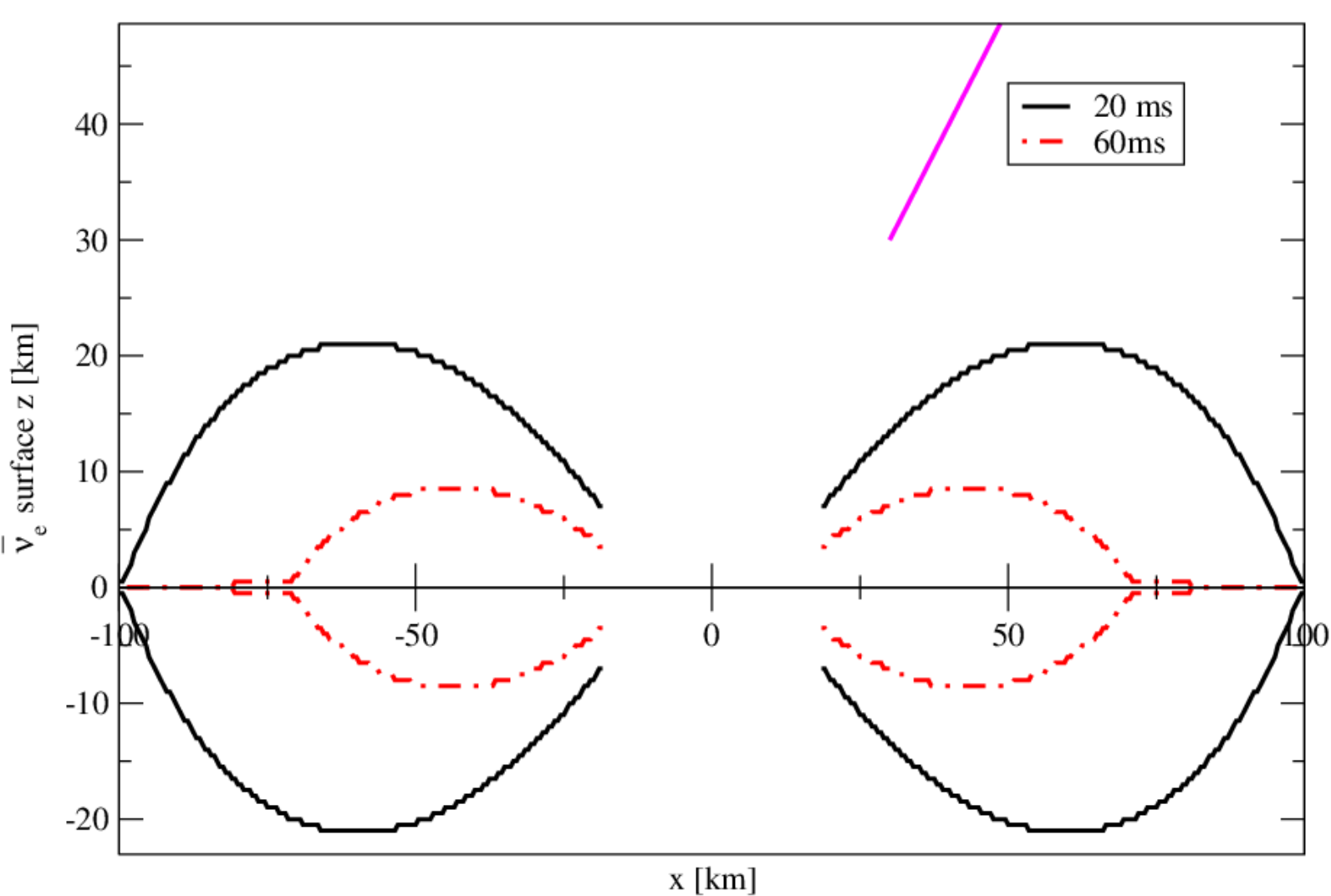}
\caption{A transversal cut of the electron antineutrino surfaces at two different times, t=20 and 60 ms.
The magenta line shows the outflow trajectory starting at x=30 km, z=30 km, It is from this point that
we launch the outflow and follow the evolution of the neutrino reaction rates.}
\label{fig-2}       
\end{figure}

For the outflow we adopt standard neutrino-driven wind trajectories parameterized in entropy and timescale or
acceleration. See for example 
ref. \citep{Qian} for supernova and ref. \citep{Surman:2004sy} for
accretion disk outflows. Our outflow follows a radial streamline that starts at $r=30$ km 
from the black hole and extends to thousands of kilometers away.  
Figure \ref{fig-2}
shows the electron antineutrino surfaces at two different times t=20 and 60 ms, 
and a segment of the outflow trajectory (magenta line). As time passes
and material is dragged into the black hole the neutrino surfaces shrink. 

\subsection{Neutrino fluxes}
We calculate the neutrino fluxes for this astrophysical environment, as in eq. \ref{observedflux}.
The observers are the points of the outflow trajectory where the reactions of eqs.
\ref{eq:capa} and \ref{eq:capb} take place. The emitters are the points on the neutrino surfaces.

For each point in the outflow we find the null geodesics that connect them to each point on
the neutrino surface. This procedure determines the solid angles $d\Omega_{ob}$. The redshift $1+z$ also
depends on the position of the outflow trajectory $r_{ob}$ and on the emission points 
on the neutrino surface $r_{em}$. We calculate both redshifts and null geodesics in the
Schwarzschild metric in a similar way as in ref. \cite{GRCaballero}.

\begin{figure}
\centering
\includegraphics[width=\linewidth,trim= 0mm 0mm 0mm 100mm, clip]{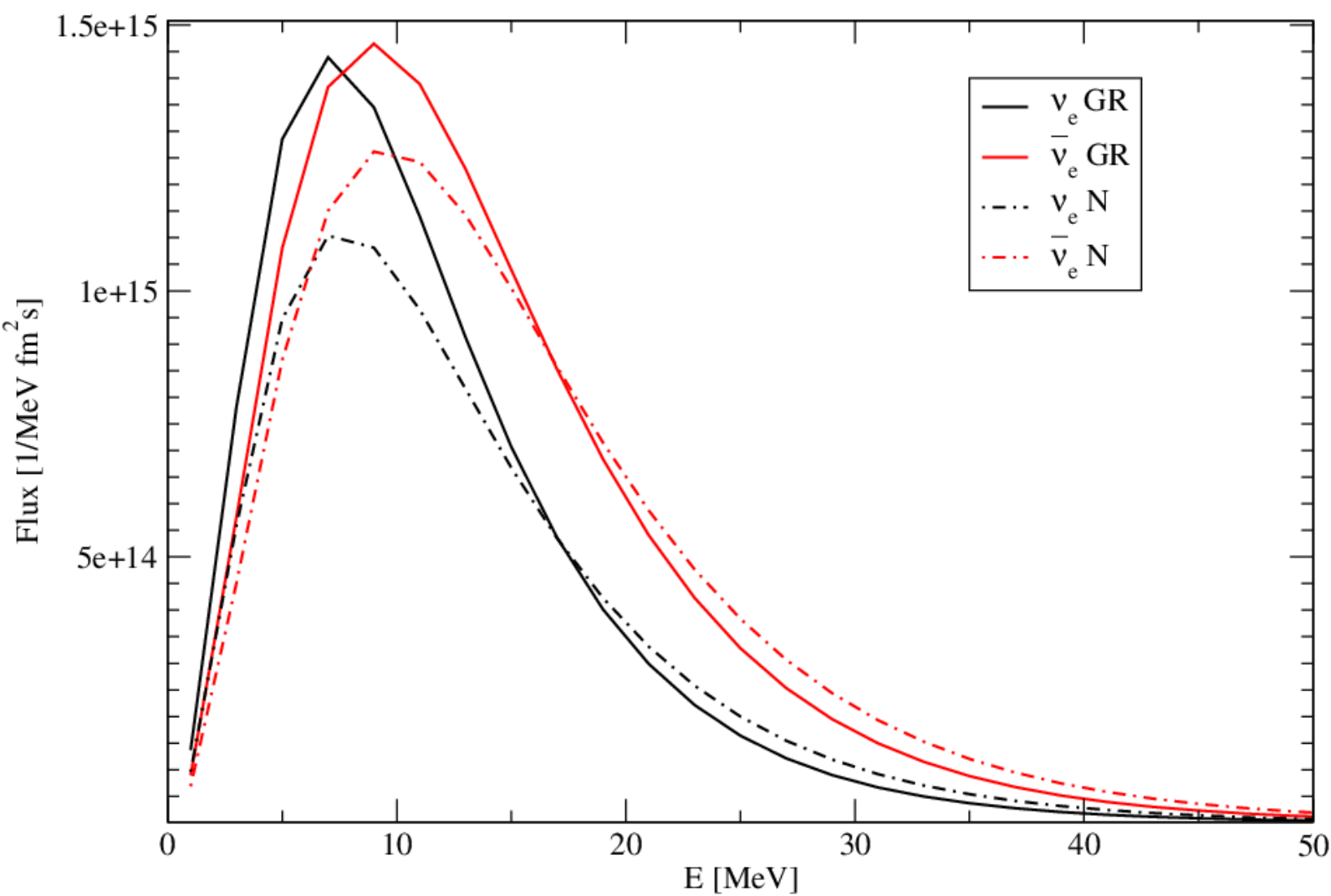}
\caption{Electron neutrino (black) and antineutrino (red) fluxes as registered at 20 ms at a point located 
at x=z=100 km from the center 
of the black hole. 
The solid lines correspond to a general relativistic calculation of the fluxes while 
the dotted-dashed lines describe fluxes in a Newtownian calculation.}
\label{fig-3}       
\end{figure}

At $t =20$ ms electron antineutrinos are hotter (see colored scale in figure \ref{fig-1}) and their fluxes
are larger than electron neutrino fluxes. This is true regardless of the space-time curvature, 
as can been seen 
when we compare the fluxes by flavor in figure \ref{fig-3} (black vs red lines), where we have plotted the fluxes seen at $z=x=100$ km.
This difference result in more antineutrino captures
on protons, driving the material neutron rich. 
When general relativistic effects are included 
the more energetic antineutrinos that are emitted closer to the black hole are more redshifted than the neutrinos. This causes
the large energy tails of the fluxes to be reduced (compare dotted-dashed lines for Newtonian (N) neutrinos 
with solid lines for the relativistic ones (GR) in figure \ref{fig-3}).
As a result in a curved space-time the 
electron antineutrinos capture rates are more reduced when compared to the Newtonian rates
and the material becomes less neutron rich. 
As an example for an outflow trajectory with an entropy per baryon of 30, the
Newtonian electron fraction $\tilde Y_e$ tends to 0.47 near $z=x=100$ km, while the relativistic 
is $\tilde Y_e=0.49$ (for an illustration of the behavior of electron fraction,
with a specific outflow, with and without general relativity see figure 2 of ref. \cite{ni56}).

\begin{figure}
\centering
\includegraphics[width=\linewidth,trim= 0mm 0mm 0mm 100mm, clip]{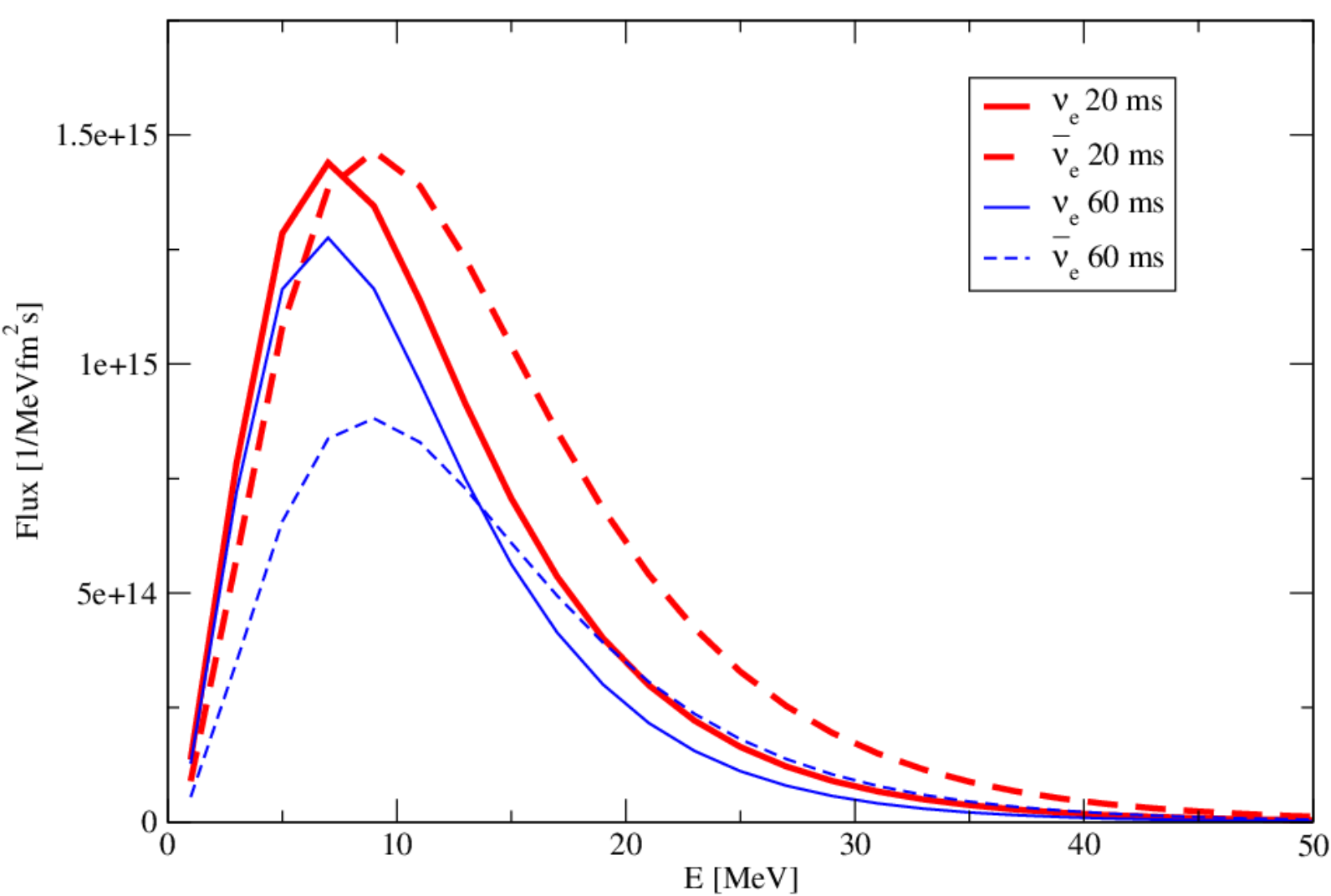}
\caption{Comparison of electron neutrino (solid) and antineutrino (dashed) fluxes registered at x=z=100 km. 
Red lines correspond to fluxes at t=20 ms and blue lines to t=60 ms. }
\label{fig-4}       
\end{figure}

Note however, that the dynamical evolution changes the emission points
as is shown in figure
\ref{fig-2}. Then the time dependence of the neutrino surfaces plays an important role. Figure \ref{fig-4}
shows the neutrino fluxes at two different times $t=20$ ms (red lines) and $t=60$ ms (blue lines) as
seen by an observer located at x$_{ob}$= z$_{ob}=100$ km from the black hole. At $t=20$ ms
electron antineutrino fluxes are larger than the electron neutrino fluxes, as it was dicussed above
(compare solid vs dashed red lines). However,
we see the opposite behavior at $t=60$ ms (see blue lines in figure \ref{fig-4}): the electron neutrino fluxes are larger. This is because as time passes
more material has been dragged into the black hole, and althought both surfaces shrink, the electron
antineutrino one is much more reduced. As a result at $t=60$ ms, neutrino capture on neutrons (eq. \ref{eq:capb})
dominates
and the material becomes proton rich. When general relativity is taken into account the reduction of
the antineutrino fluxes is stronger because the antineutrinos are even more redshifted
than neutrinos. This makes the medium even more
proton rich. In the example mentioned above this translates into
a Newtonian $\tilde Y_e \approx 0.53$, while with GR $\tilde Y_e \approx 0.56$ (this can be seen in the magenta lines of figure in ref \cite{ni56}).

\section{Production of $^{56}$Ni}
The dynamical study of the final abundances requires the knowledge of the electron fraction
as a function of time.
So far, we have shown the influence of gravity on the electron fraction at
two different times, $t=20$ and $t=60$ ms.
A detailed calculation of the electron fraction for all times would require repeating the steps described
above for small time steps in this interval. This would include finding the neutrino surfaces and the null geodesics
at all times and for every point of the outflow trajectory. This procedure would be computationally
expensive. Instead we can think of a simple model. It is natural to expect that the electron fraction
would lie in between the limiting 
values obtained at the two snapshops
$t=20$ and $t=60$ ms mentioned above $\tilde Y_e=0.49$ and $\tilde Y_e=0.56$.
In ref \cite{ni56} we proposed at linear time dependency of the
reaction rates to emulate this time evolution. 

\begin{figure}
\centering
\includegraphics[width=\linewidth,trim= 0mm 0mm 0mm 100mm, clip]{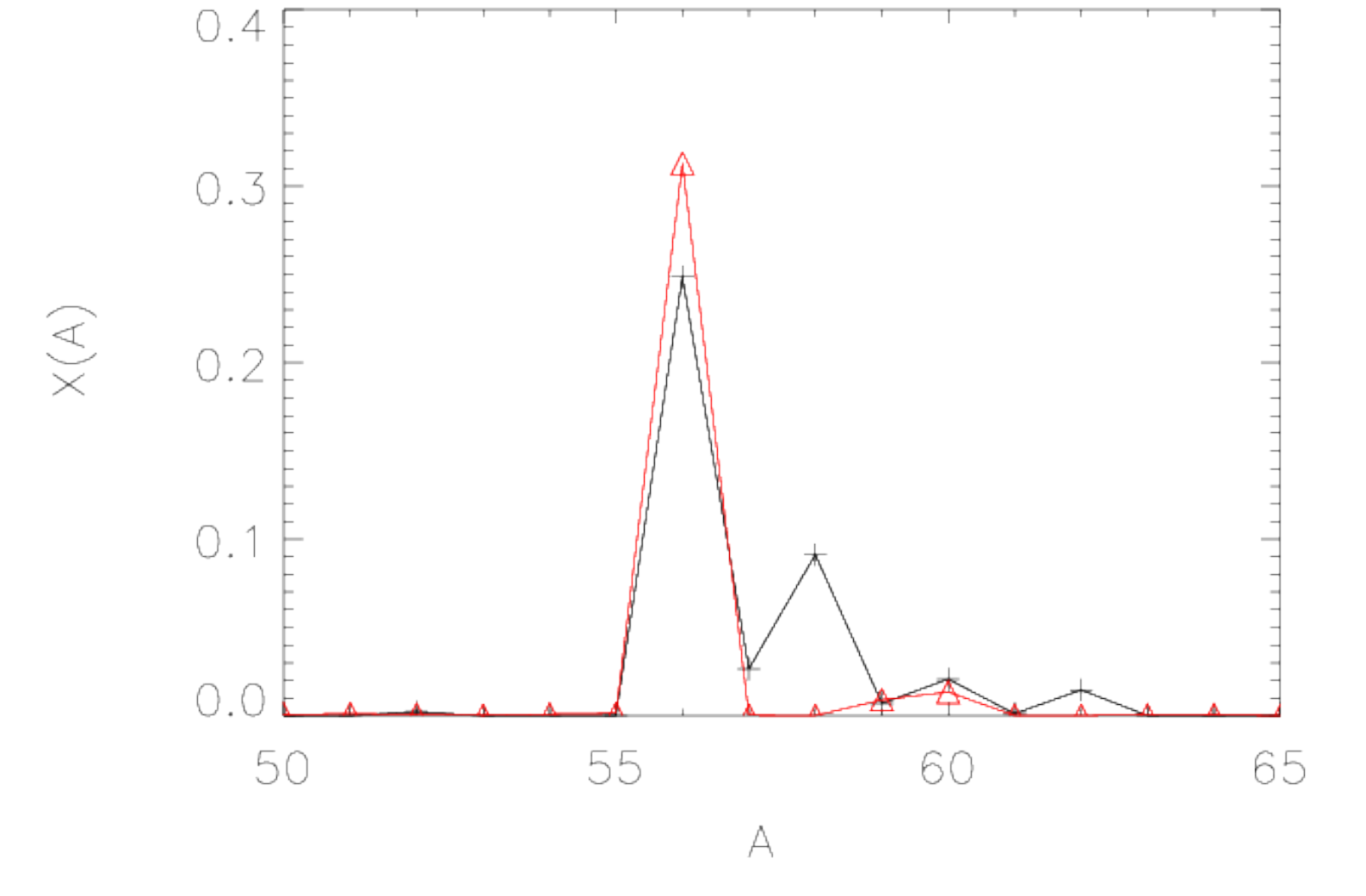}
\caption{Final mass fractions for an outflow trajectory with $s/k=30$ and $\tau=20$ ms, corresponding to a calculation
with (red) and without (black) general relativistic corrections.
 }
\label{fig-5}       
\end{figure}

In our nucleosynthesis calculations of ref \cite{ni56} we also sample the outflow parameter space,
allowing a wide variety of thermodynamical outflow conditions.
Entropies per baryon $s/k$ were
allowed to take values from 20 to 80 , and the effective dynamic
time scale $\tau$ was varied beteween 10 and 100 ms. 
A description of the nuclear reaction network  can be found in refs. \cite{Surman:2011,Hix:1999, McLaughlin:1997qi}. 
Under the conditions described above
we performed nucleosythesis calculations to find the abundances produced in these outflows.
In figure \ref{fig-5} we show the final mass fraction $X(A)$ vs mass number $A$, for an outflow
trajectory with entropy per baryon $s/k=30$ and dynamic timescale $\tau=20$ ms. The red lines
show the GR mass fractions while the black lines correspond to a Newtonian calculation. The 
enhacement in the production of $^{56}$Ni in the GR mass fractions is the result of the larger electron
fraction for this case.

For the majority of the outflow conditions we find proton-rich nuclei, with $^{56}$Ni the most abundant. 
This is a direct consequence of the time evolution
of the neutrino fluxes. The GR corrections further enhance the proton-richness of the outflow,
by increasing the electron fraction. Note however, that 
for higher entropies, the simulations with GR corrections would lead
to $Y_e$ above the optimum range for a large $^{56}$Ni production. Therefore the
mass fraction of $^{56}$Ni would be larger in the
Newtonian case (see \cite{ni56} for
details on the $^{56}$Ni abundance fractions as a function of the outflow condtions). 
\section{Conclusions}
The flux of neutrinos observed from a source with a massive central object  
is significantly different from the flux emitted from the same source in a graviational field free space.
This difference affects the neutrino absorption rates on nucleons and therefore the electron fraction of the
medium. The changes introduced in the electron fraction by the gravitaional field are important enough
to alter the nucleosynthesis final abundances.

Due to the initially low electron fraction of their progenitors, merger-type accretion disks have been considered
to be good candidates for the synthesis of heavy neutron rich nuclei. 
We study the synthesis of elements that occur in black hole accretion disk 
outflows. Our disk model is based 
on a hydrodynamical simulation, and the outflow model is similar to standard supernova neutrino winds.
We find that gravity plays an important role in setting the electron fraction via its influence in  
the behavior of neutrinos.
The over all change in the neutrino fluxes is a reduction of the high energy tails due to redshifts.
This effect is stronger in the electron antineutrino channel, driving the material proton-rich.
We find that time evolution plays an important role as the neutrino surfaces shrink when matter 
is dragged into the black
hole. This reduction in the neutrino surfaces combined with the
stronger redshifts leads to even more proton rich material and the synthesis of larger amounts of $^{56}$Ni
for a wide range of outflow conditions.

\begin{acknowledgement}
This work was partially supported by the Natural Sciences and
Engineering Research Council of Canada (NSERC)(OLC), the Helmholtz Alliance Program of the Helmholtz Association, contract HA216/EMMI 
``Extremes of Density and Temperature: Cosmic Matter in the Laboratory'' (OLC), and the Department of Energy under contracts DE-FG02-
05ER41398 (RS) and DE-FG02-02ER41216 (GCM).
\end{acknowledgement}

\end{document}